\documentclass[aps,prd,superscriptaddress,showpacs,tighten,nofootinbib,twocolumn]{revtex4}
\usepackage{amsfonts}
\usepackage{amssymb}
\usepackage{amsmath,color}
\usepackage{epsfig}
\usepackage{graphicx,epsf,epsfig}
\usepackage{bbm}
\def\slc#1{\setbox0=\hbox{$#1$}           
    \dimen0=\wd0                                 
    \setbox1=\hbox{/} \dimen1=\wd1               
    \ifdim\dimen0>\dimen1                        
       \rlap{\hbox to \dimen0{\hfil/\hfil}}      
       #1                                        
    \else                                        
       \rlap{\hbox to \dimen1{\hfil$#1$\hfil}}   
       /                                         
    \fi}

\begin{document}
\vspace*{-2cm}
\begin{flushright}
MPP-2009-210
\end{flushright}
\title{The Minimal Seesaw Model at the TeV Scale}
\author{He Zhang} \email{zhanghe@kth.se}

\affiliation{Department of Theoretical Physics, School of
Engineering Sciences, Royal Institute of Technology (KTH) --
AlbaNova University Center, Roslagstullsbacken 21, 106 91 Stockholm,
Sweden}

\author{Shun Zhou}
\email{zhoush@mppmu.mpg.de}

\affiliation{Max-Planck-Institut f{\"u}r Physik
(Werner-Heisenberg-Institut), F{\"o}hringer Ring 6, 80805
M{\"u}nchen, Germany}
\begin{abstract}
We point out that the minimal seesaw model can provide a natural
framework to accommodate tiny neutrino masses, while its
experimental testability and notable predictiveness are still
maintained. This possibility is based on the observation that two
heavy right-handed Majorana neutrinos in the minimal seesaw model
may naturally emerge as a pseudo-Dirac fermion. In a specific
scenario, we show that the tri-bimaximal neutrino mixing can be
produced, and only the inverted neutrino mass hierarchy is allowed.
The low-energy phenomena, including non-unitarity effects in
neutrino oscillations, neutrinoless double-beta decays and rare
lepton-flavor-violating decays of charged leptons $\ell^{}_\alpha
\to \ell^{}_\beta \gamma$, have been explored. The collider
signatures of the heavy singlet neutrino are also briefly discussed.
\end{abstract}
\maketitle
\section{Introduction} \label{sec:intro}

Recently, a lot of attention has been focused on the experimental
testability of seesaw models for neutrino masses at the CERN Large
Hadron Collider (LHC). In the typical extension of the standard
model (SM), three right-handed neutrinos are introduced and assigned
large Majorana masses
\cite{Minkowski:1977sc,Yanagida:1979as,GellMann:1980vs,Mohapatra:1979ia}.
In this case, the SM neutrinos acquire tiny Majorana masses via the
type-I seesaw mechanism, i.e. the effective mass matrix of light
neutrinos is given by $M^{}_\nu \approx - M^{}_{\rm D} M^{-1}_{\rm
R} M^T_{\rm D}$. Here the Dirac mass is naturally around the
electroweak scale $M^{}_{\rm D} \sim \Lambda^{}_{\rm EW} \equiv
100~{\rm GeV}$ and the heavy Majorana neutrino masses are extremely
large $M^{}_{\rm R} \sim 10^{10\cdots14}~{\rm GeV}$. However, this
type-I seesaw model suffers from the lack of testability, because
right-handed neutrinos are too heavy to be produced in current
collider experiments.

In order for the type-I seesaw model to be testable, we have to
implement the structural cancellation condition $M^{}_{\rm D}
M^{-1}_{\rm R} M^T_{\rm D} \approx {\bf 0}$, which can lead to
sub-eV neutrino masses but keep heavy Majorana neutrino masses as
low as several hundred GeV \cite{Pilaftsis:1991ug,Kersten:2007vk}.
However, the fine-tunings of the structures of $M^{}_{\rm D}$ and
$M^{}_{\rm R}$ are unavoidable, which seems unnatural. It has been
pointed out that the heavy pseudo-Dirac neutrinos may account for
tiny neutrino masses, such as in the so-called inverse seesaw
models, which can be tested in collider experiments, and are free of
the fine-tuning problem \cite{Mohapatra:1986bd}. Nevertheless, more
singlet fermions should be added in order to form pseudo-Dirac
fermions together with singlet right-handed neutrinos. For instance,
in the minimal version of inverse seesaw model, we must have four
singlet fermions to guarantee two massive light neutrinos
\cite{Malinsky:2009df}. Even more singlet fermions are required in
the realization of multiple seesaw mechanisms \cite{Xing:2009hx},
which are the direct generations of the type-I and the inverse
seesaw models.

We propose that the minimal type-I seesaw model (MSM) with only two
right-handed neutrinos may be the most natural candidate for
realistic and testable neutrino mass models at the TeV scale. The
reason is simply that these two right-handed neutrinos themselves
can be combined together to make a Dirac fermion in the $U(1)$
symmetry limit. They may also be embedded into a two-dimensional
representation of the discrete flavor symmetry groups, such as the
permutation group $S_4$ \cite{Yamanaka:1981pa,Zhang:2006fv}. The
soft symmetry-breaking terms then give rise to tiny neutrino masses,
while the heavy pseudo-Dirac neutrino provides us with rich
low-energy phenomena, e.g. the non-unitarity effects in neutrino
oscillations and the lepton-flavor-violating decays of charged
leptons. Furthermore, the tri-lepton signals $pp \to
\ell_{\alpha}^{\pm} \ell_{\beta}^{\pm}\ell_{\gamma}^{\mp}
\nu(\bar\nu)+\text{jets}$ of the pseudo-Dirac neutrino can be
discovered at the LHC. Due to the minimal number of model
parameters, the observables at low energies and in the collider
experiments are intimately correlated with each other, and then
serve as a cross test of our minimal TeV seesaw model.

The remaining part of this work is organized as follows. In
Sec.~\ref{sec:model}, we first present the structure of our model.
The neutrino mass spectra and neutrino mixing patterns are discussed
in detail in Sec.~\ref{sec:mixing}. The implications for low-energy
phenomena and possible collider signatures are addressed in
Sec.~\ref{sec:LHC}. Finally, a brief summary is given in
Sec.~\ref{sec:summary}.

\section{The model}\label{sec:model}

In the minimal seesaw model \cite{Frampton:2002qc,Guo:2006qa}, we
extend the SM by introducing two heavy right-handed neutrinos, which
are singlets under the SM gauge group. The Lagrangian relevant for
neutrino masses is
\begin{eqnarray}\label{equ:L}
-{\cal L}_{\rm mass} = \overline{\nu^{}_{\rm L}} M^{}_{\rm D}
n^{}_{\rm R} + \frac{1}{2} \overline{n^c_{\rm R}} M^{}_{\rm R}
n^{}_{\rm R} + {\rm h.c.} \, ,
\end{eqnarray}
where $\nu^{}_{\rm L} = (\nu^{}_{e{\rm L}}, \nu^{}_{\mu {\rm
L}},\nu^{}_{\tau {\rm L}})^T$ and $n^{}_{\rm R} = (n^{}_{1\rm R},
n^{}_{2\rm R})^T$ stand for left- and right-handed neutrinos,
respectively. To effectively suppress the light neutrino masses
while keeping heavy ones around the TeV scale, we additionally
impose a global $U(1)$ symmetry on the generic Lagrangian, under
which the charges of the SM lepton doublets are opposite to
$n^{}_{2\rm R}$ but equal to $n^{}_{1\rm R}$. With the help of such
a lepton-number-like symmetry, the mass matrices $M^{}_{\rm D}$ and
$M^{}_{\rm R}$ in Eq.~\eqref{equ:L} take very simple forms
\begin{eqnarray}\label{equ:mDR}
M^{}_{\rm D} = v \left(\begin{matrix} y^{}_e & y^{}_\mu & y^{}_\tau
\cr 0 & 0 & 0
\end{matrix}\right)^T \, ,~~~ M^{}_{\rm R} = \left(\begin{matrix} 0 & M
\cr M & 0 \end{matrix}\right)
\end{eqnarray}
with $v\simeq174~{\rm GeV}$ being the vacuum expectation value of
the Higgs field. In the flavor basis ($\nu^{}_{\rm L}, n^c_{\rm
R}$), the overall $5\times5$ neutrino mass matrix reads
\begin{eqnarray}\label{eq:Mnu}
{\cal M}^{}_{\nu}=\left(\begin{array}{cc}
0 & M^{}_{\rm D} \\
M^T_{\rm D} & M^{}_{\rm R}
\end{array}\right) \, .
\end{eqnarray}
Note that the rank of ${\cal M}^{}_\nu$ is two, so three light
neutrinos are massless and two heavy Majorana neutrinos are
degenerate in mass, as a consequence of the additional global
symmetry. This can be easily verified by noting that the mass
matrices in Eq.~\eqref{equ:mDR} satisfy the relation $M^{}_{\rm D}
M^{-1}_{\rm R} M^T_{\rm D} = {\bf 0}$, which implies that light
neutrino masses are vanishing to all orders. Therefore, the masses
of heavy right-handed neutrinos have nothing to do with light
neutrinos and can be located at a relatively low scale, e.g., around
the TeV scale. On the other hand, the symmetric matrix $M^{}_{\rm
R}$ can be diagonalized via an orthogonal transformation $V^T_{\rm
R} M^{}_{\rm R} V^{}_{\rm R} = {\rm Diag}\{-M, M\}$ with
\begin{eqnarray}\label{eq:VR}
V^{}_{\rm R} = \frac{1}{\sqrt{2}}\left(\begin{matrix} 1 & 1 \cr -1 &
1 \end{matrix}\right) \, .
\end{eqnarray}
Thus the mass eigenstates of heavy right-handed neutrinos $P^{}_1$
and $P^{}_2$ possess identical masses but opposite CP parities, and
they constitute a four-component Dirac particle $P=(P^{}_1 +
P^{}_2)/\sqrt{2}$ with mass being $M$ \cite{Bilenky:1987ty}.

In order to accommodate light neutrino masses, one can add small
perturbations to ${\cal M}^{}_\nu$, which softly break the global
$U(1)$ symmetry. There are in principle four classes of
soft-breaking perturbations to ${\cal M}^{}_\nu$, and the most
general form of the perturbed neutrino mass matrix is given by
\begin{eqnarray}\label{equ:MnuP}
{\cal M}^{}_\nu = \left(\begin{matrix} \kappa_{ee} & \kappa_{e\mu} &
\kappa_{e\tau} & v y_e & \varepsilon_e \cr \kappa_{\mu e} &
\kappa_{\mu\mu} & \kappa_{\mu\tau} & v y_\mu & \varepsilon_\mu \cr
\kappa_{\tau e} & \kappa_{\tau\mu} & \kappa_{\tau\tau} & v y_\tau &
\varepsilon_\tau \cr v y_e & v y_\mu & v y_\tau & \mu^\prime & M \cr
\varepsilon_e & \varepsilon_\mu & \varepsilon_\tau & M & \mu
\end{matrix}\right) \, .
\end{eqnarray}
All the above perturbation terms $\kappa^{}_{\alpha \beta}$,
$\varepsilon^{}_\alpha$, $\mu^\prime$ and $\mu$ (for $\alpha, \beta
= e, \mu, \tau$) break the lepton number conservation, and hence
bring in neutrino masses proportional to the corresponding
couplings. Some comments on the possible origins of these
perturbations are in order:

(i) The $\kappa$ term corresponds to a purely Majorana mass term of
light neutrinos, which can be realized in a more complicated theory
with additional contributions to neutrino masses. A typical example
is the type-(I+II) seesaw model
\cite{Schechter:1980gr,Lazarides:1980nt,Mohapatra:1980yp}, where an
$SU(2)$ triplet Higgs with mass much larger than the electroweak
scale is involved. At lower-energy scales, the decoupling of the
triplet Higgs will result in light neutrino masses together with
non-standard interactions through the tree-level exchange of the
neutral scalar \cite{Malinsky:2008qn}. Another possibility is to
incorporate extra SM singlet or triplet fermions, which give birth
to a neutrino mass operator similar to that in the type-I or
type-III seesaw models \cite{Foot:1988aq}. Although feasible, the
above mechanisms are always pestered with too many parameters, which
render the models neither predictive nor economical.

(ii) The $\mu$ term in Eq.~\eqref{equ:MnuP} is a bare Majorana mass
insertion violating the lepton number by two units, which is also
realized in the inverse seesaw framework
\cite{Mohapatra:1986bd,Malinsky:2009gw}. In the presence of the
$\mu$ term, the light neutrino mass matrix can be obtained from the
inverse seesaw formula
\begin{eqnarray}\label{equ:Minv}
M^{}_\nu \simeq \frac{\mu}{M^2} M^{}_{\rm D} M^T_{\rm D} \, .
\end{eqnarray}
Hence the smallness of neutrino masses is attributed to both the
small $\mu$ parameter and the ratio $M^{}_{\rm D}/M$. However, as
pointed out in Ref.~\cite{Malinsky:2009df}, at least two pairs of
singlet heavy neutrinos are required in order to enhance the rank of
${\cal M}^{}_\nu$ from two to four. One can also see this point from
Eq.~\eqref{equ:Minv} that the rank of $M^{}_\nu$ is exactly one,
which definitely comes into conflict with the observed two
mass-squared differences in neutrino oscillation experiments.

(iii) The $\mu^\prime$ term in Eq.~\eqref{equ:MnuP} does not
contribute to neutrino masses at the tree level. However, it may
radiatively generate neutrino masses via one-loop diagrams involving
right-handed neutrinos and gauge bosons \cite{Pilaftsis:1991ug}. In
addition, due to the corrections induced by the $\mu^\prime$ term,
the masses of $P^{}_1$ and $P^{}_2$ are not exactly equal, and thus
the small mass splitting between heavy neutrinos could naturally
make the resonant leptogenesis mechanism feasible
\cite{Pilaftsis:2005rv}. In analogy with the $\mu$-term corrections,
the drawback is that only one light neutrino may acquire mass, and
hence the $\mu^\prime$ term is not phenomenologically adequate.

(iv) The $\varepsilon$ term softly violates the extra $U(1)$
symmetry but enhances the rank of ${\cal M}^{}_\nu$ from two to
four, which is required by the neutrino oscillations. Such a
coupling could be easily realized in grand unified theories, e.g. in
the supersymmetric $SO(10)$ model with a very low $B-L$ scale
\cite{Malinsky:2005bi}. As we will show later in this case, neutrino
masses are naturally tiny, since they are proportional to
$\varepsilon$, and are further suppressed by the mass ratio
$M^{}_{\rm D}/M$. In the following, we will only concentrate on this
particularly interesting pattern of neutrino mass generation, and
figure out the phenomenological consequences in detail. The overall
neutrino matrix can be obtained from Eq.~\eqref{equ:MnuP} by setting
all $\kappa$, $\mu$ and $\mu^\prime$ to zero.

Without loss of generality, one can always redefine the lepton
fields so as to remove the corresponding phases of $y^{}_\alpha$ and
$M$, leaving only $\varepsilon_\alpha$ complex. Therefore, we shall
assume $y^{}_\alpha$ and $M$ to be real throughout the following
discussions. The total neutrino matrix ${\cal M}^{}_\nu$ can be
diagonalized by the unitary transformation
\begin{eqnarray}\label{eq:diag}
V^{\dagger} {\cal M}^{}_{\nu} V^{*}= \widehat{\cal M}^{}_\nu \equiv
{\rm Diag}\{m^{}_1, m^{}_2, m^{}_3, -M, M\} \ ,
\end{eqnarray}
where $V$ is a $5\times 5$ unitary matrix, and $m^{}_i$ (for
$i=1,2,3$) are masses of three light neutrinos. In the leading-order
approximaiton, the effective mass matrix of light neutrinos is given
by the type-I seesaw formula
\begin{eqnarray}\label{eq:mnu}
M^{}_\nu \simeq - M^{}_{\rm D} M^{-1} M^T_{\rm D} = - \varepsilon
F^T- F \varepsilon^T \, ,
\end{eqnarray}
where $\varepsilon = (\varepsilon^{}_e, \varepsilon^{}_\mu,
\varepsilon^{}_\tau)^T$ and $F = \omega(y^{}_e, y^{}_\mu,
y^{}_\tau)^T$ with $\omega \equiv v/M$. In general, $M^{}_\nu$ can
be diagonalized by a $3\times 3$ unitary matrix as $U^\dagger
M^{}_{\nu} U^* = \widehat{M}^{}_\nu  \equiv {\rm Diag} \{m^{}_1,
m^{}_2, m^{}_3\}$, and $U$ is usually parametrized in the standard
form
\begin{eqnarray}\label{eq:para}
U  = P^{}_{\phi} R^{}_{23}(\theta^{}_{23}) P^{}_{\delta}
R^{}_{13}(\theta^{}_{13}) P^{-1}_{\delta} R^{}_{12}(\theta^{}_{12})
P^{}_{\rm M} \ ,
\end{eqnarray}
where $R_{ij}$ correspond to the elementary rotations in the
$ij=23$, $13$, and $12$ planes, $\theta^{}_{ij}$ are the rotation
angles, while $P^{}_{\delta}={\rm Diag}\{1,1,{\rm e}^{{\rm
i}\delta}\}$ and $P^{}_{\rm M}={\rm Diag}\{1, {\rm e}^{{\rm
i}\rho},1\}$ contain the Dirac and Majorana CP-violating phases,
respectively. Note that only one Majorana phase is needed to
parametrize $U$ since one light neutrino is massless, which is the
salient feature of MSM. The phases in $P_\phi = {\rm Diag}(e^{{\rm
i}\phi_1},e^{{\rm i}\phi_2},e^{{\rm i}\phi_3})$ are usually rotated
away in the SM context but must be kept in the current model. It
should be noticed that $U$ is not exactly the matrix governing
neutrino oscillations, even if we choose a basis where the
charged-lepton mass matrix is diagonal. To clarify this point, we
turn back to the $5\times 5$ unitary matrix $V$ in
Eq.~\eqref{eq:diag}, which can be rewritten in a block form
\begin{eqnarray}
V=\left(\begin{matrix} N_{3\times3} & R_{3\times2} \cr  S_{2\times3}
& T_{2\times 2}
\end{matrix} \right) \, .
\end{eqnarray}
The approximate expression of each block can be found in Ref.
\cite{Schechter:1981cv}, and the relevant ones are
\begin{eqnarray} \label{eq:N}
N \simeq \left(1-\frac{1}{2}FF^\dagger \right) U \,, ~~~~ R \simeq Q
V^{}_{\rm R} \, ,
\end{eqnarray}
with $Q\equiv ({\bf 0}, F)$ being a $3\times 2$ matrix. The flavor
eigenstates of neutrinos are then the superpositions of light
neutrino mass eigenstates $\nu^{}_{m{\rm L}}=(\nu^{}_{1{\rm
L}},\nu^{}_{2{\rm L}},\nu^{}_{3{\rm L}})$ and the heavy one $P$.
More specifically, the flavor eigenstates of light neutrinos can be
expressed as $\nu^{}_{\rm L} \simeq N \nu^{}_{m{\rm L}} + F P^c$,
which indicates that $P^c$ mixes with left-handed neutrinos and
enters into the weak interactions after the electroweak symmetry
breaking. Therefore, the leptonic charged-current interactions in
the mass basis read
\begin{eqnarray}\label{eq:Lcc}
{\cal L}_{\rm CC} & = & -\frac{g}{\sqrt{2}}   \overline{\ell_{\rm
L}} \gamma^\mu \left( N  \nu^{}_{m{\rm L}} + F P^c \right)W^-_\mu +
{\rm h.c.} \, .
\end{eqnarray}
It is obvious that the neutrino mixing matrix $N$ appearing in the
charged current is non-unitary. As we shall show, the mixing between
light and heavy neutrinos will bring in several significant
phenomenological consequences, in particular, when the scale of the
heavy neutrino masses is accessible to future colliders.

\section{neutrino masses and mixing}
\label{sec:mixing}

First, we should consider the neutrino mass spectra and flavor
mixing matrix. It is straightforward to obtain the light neutrino
mass matrix from Eq.~\eqref{eq:mnu}
\begin{eqnarray}\label{eq:mnu1}
M^{}_\nu = \omega \left(\begin{matrix}2 \varepsilon^{}_e y^{}_e &
\varepsilon^{}_e y^{}_\mu + \varepsilon^{}_\mu y^{}_e &
\varepsilon^{}_e y^{}_\tau + \varepsilon^{}_\tau y^{}_e \cr
\varepsilon^{}_e y^{}_\mu + \varepsilon^{}_\mu y^{}_e & 2
\varepsilon^{}_\mu y^{}_\mu & \varepsilon^{}_\mu y^{}_\tau +
\varepsilon^{}_\tau y^{}_\mu \cr \varepsilon^{}_e y^{}_\tau +
\varepsilon^{}_\tau y^{}_e & \varepsilon^{}_\mu y^{}_\tau +
\varepsilon^{}_\tau y^{}_\mu & 2 \varepsilon^{}_\tau y^{}_\tau
\end{matrix}\right) \, ,~~
\end{eqnarray}
where the irrelevant minus sign has been omitted. Since the
experimental data on neutrino oscillations suggest the tri-bimaximal
mixing pattern \cite{Harrison:2002er,Xing:2002sw}, a $\mu$-$\tau$
symmetry is particularly favorable in constructing the neutrino mass
matrix. To this end, we assume here that the relations
$\varepsilon_\mu=\varepsilon_\tau$ and $y_\mu=y_\tau$ hold. Defining
$A=2\omega \varepsilon^{}_e y^{}_e$, $B=\omega(\varepsilon^{}_e
y^{}_\mu + \varepsilon^{}_\mu y^{}_e)$ and $C=2\omega
\varepsilon^{}_\mu y^{}_\mu$, one can rewrite Eq.~\eqref{eq:mnu1} as
follows
\begin{eqnarray}\label{eq:mnu2}
M^{}_\nu =\left(\begin{matrix} A & B & B \cr B & C & C \cr B & C & C
\end{matrix}\right) \, ,
\end{eqnarray}
which can be further put into a $2\times2$ block form by a maximal
rotation
\begin{eqnarray}\label{eq:mnu3}
M^\prime_\nu = R^{}_{23}(\frac{\pi}{4}) M^{}_\nu R^T_{23}
(\frac{\pi}{4}) = \left(\begin{matrix} A & \sqrt{2}B & 0 \cr
\sqrt{2}B & 2C & 0 \cr 0 & 0 & 0
\end{matrix}\right) \, .
\end{eqnarray}
We have found that only the inverted mass hierarchy $m^{}_2
> m^{}_1 > m^{}_3 = 0$ is compatible with a maximal mixing pattern
in the $2\leftrightarrow 3$ sector. The left-up block matrix in Eq.
\eqref{eq:mnu3}, denoted as $\tilde{M}^\prime_\nu$, is a general
$2\times 2$ symmetric matrix, which can be diagonalized as
$U_0^\dagger \tilde{M}^\prime_\nu U^*_0= {\rm Diag}\{m^{}_1,
m^{}_2\}$ with
\begin{eqnarray}\label{eq:U0}
U^{}_0 = \left(\begin{matrix} c^{}_\theta & s^{}_\theta e^{-{\rm
i}\phi} \cr -s^{}_\theta e^{{\rm i}\phi} & c^{}_\theta
\end{matrix}\right) \left(\begin{matrix}  e^{{\rm i}\psi_1} & 0
\cr 0 & e^{{\rm i}\psi_2}
\end{matrix}\right)  \, .
\end{eqnarray}
Here $c^{}_\theta \equiv \cos\theta$ and $s^{}_\theta \equiv
\sin\theta$ have been defined. After a lengthy but straightforward
calculation, one can obtain
\begin{eqnarray}\label{eq:theta}
\tan\phi = \frac{X}{Y} \, , ~~~~ \tan2\theta =
\frac{\sqrt{2(X^2+Y^2)}}{|A|^2 - 4 |C|^2} \; ,
\end{eqnarray}
where
\begin{eqnarray}\label{eq:XY}
X  &=&  {\rm Im}(B)  {\rm Re}(A-2C)  - {\rm Re}(B)
{\rm Im}(A-2C)  \, , \nonumber \\
Y  &=&  {\rm Im}(B)  {\rm Im}(A+2C) + {\rm Re}(B) {\rm Re}(A+2C) \,,
~~~
\end{eqnarray}
together with the mixing angles $\theta_{13}=0$ and
$\theta_{23}=45^\circ$ defined in Eq.~\eqref{eq:para}. The
non-vanishing neutrino mass eigenvalues are given by
\begin{eqnarray}\label{eq:m}
m_1 &=&  \left|A c^2_\theta - \sqrt{2}B s^2_\theta e^{-{\rm i}\phi}
+ 2C s^2_\theta e^{-2{\rm i}\phi} \right| \, ,  \nonumber \\
m_2 &=&  \left|A s^2_\theta + \sqrt{2}B s^2_\theta e^{-{\rm i}\phi}
+ 2C c^2_\theta e^{-2{\rm i}\phi} \right| \, .
\end{eqnarray}
In addition, the phase difference $\Delta\psi \equiv \psi_2 -\psi_1$
reads
\begin{eqnarray}\label{eq:psi}
\Delta\psi = \frac{1}{2}\arg\left( \frac{A s^2_\theta + \sqrt{2}B
s^2_\theta e^{-{\rm i}\phi} + 2C c^2_\theta e^{-2{\rm i}\phi} }{A
c^2_\theta - \sqrt{2}B s^2_\theta e^{-{\rm i}\phi} + 2C s^2_\theta
e^{-2{\rm i}\phi}}  \right) \, .
\end{eqnarray}
In comparison with the standard parametrization in
Eq.~\eqref{eq:para}, we can observe that $\rho = \Delta\psi - \phi$
is just the physical Majorana phase. Once the identity
$\sqrt{X^2+Y^2}=2(|A|^2-4|C|^2)$ is fulfilled, the tri-bimaximal
mixing pattern with $\theta^{}_{12} = \theta \approx 35.3^\circ$,
$\theta^{}_{23} = 45^\circ$ and $\theta^{}_{13} = 0$ is reproduced
in the leading order. The vanishing $\theta_{13}$ can be regarded as
a consequence of the exact $\mu\leftrightarrow\tau$ symmetry. If we
relax this assumption, i.e. $\varepsilon^{}_\mu \neq
\varepsilon^{}_\tau$ and (or) $y^{}_\mu \neq y^{}_\tau$, both
non-vanishing mixing angle $\theta_{13}$ and CP-violating phase
$\delta$ can be accommodated.

As we discussed above, one peculiar feature in our model is that the
mixing matrix of light neutrinos $N$ is no longer unitary. Adopting
the parametrization of non-unitary leptonic mixing matrix $N = (1 -
\eta)U$ in Ref.~\cite{Altarelli:2008yr}, we can see from Eq.
\eqref{eq:N} that
\begin{eqnarray}\label{eq:eta}
\eta & = & \frac{1}{2}FF^\dagger = \frac{\omega^2}{2}
\left(\begin{matrix} y^2_e & y^{}_e y^{}_\mu & y^{}_e y^{}_\tau \cr
y^{}_e y^{}_\mu &  y^2_\mu & y^{}_\mu y^{}_\tau \cr y^{}_e y^{}_\tau
& y^{}_\mu y^{}_\tau  &  y^2_\tau
\end{matrix}\right) \, .
\end{eqnarray}
The $\mu\leftrightarrow\tau$ symmetric feature of the effective
neutrino mass matrix $M^{}_\nu$ indicates $\eta^{}_{e\mu} \sim
\eta^{}_{e\tau}$. Since the most stringent experimental constraints
come from the lepton-flavor-violating (LFV) decay $\mu \to e\gamma$,
severe bounds on $\eta^{}_{e\mu}$ and $\eta^{}_{e\tau}$ are
expected.

\begin{figure}[t]
\begin{center}\vspace{-0.7cm}
\includegraphics[width=6.0cm,bb=30 100 680 750]{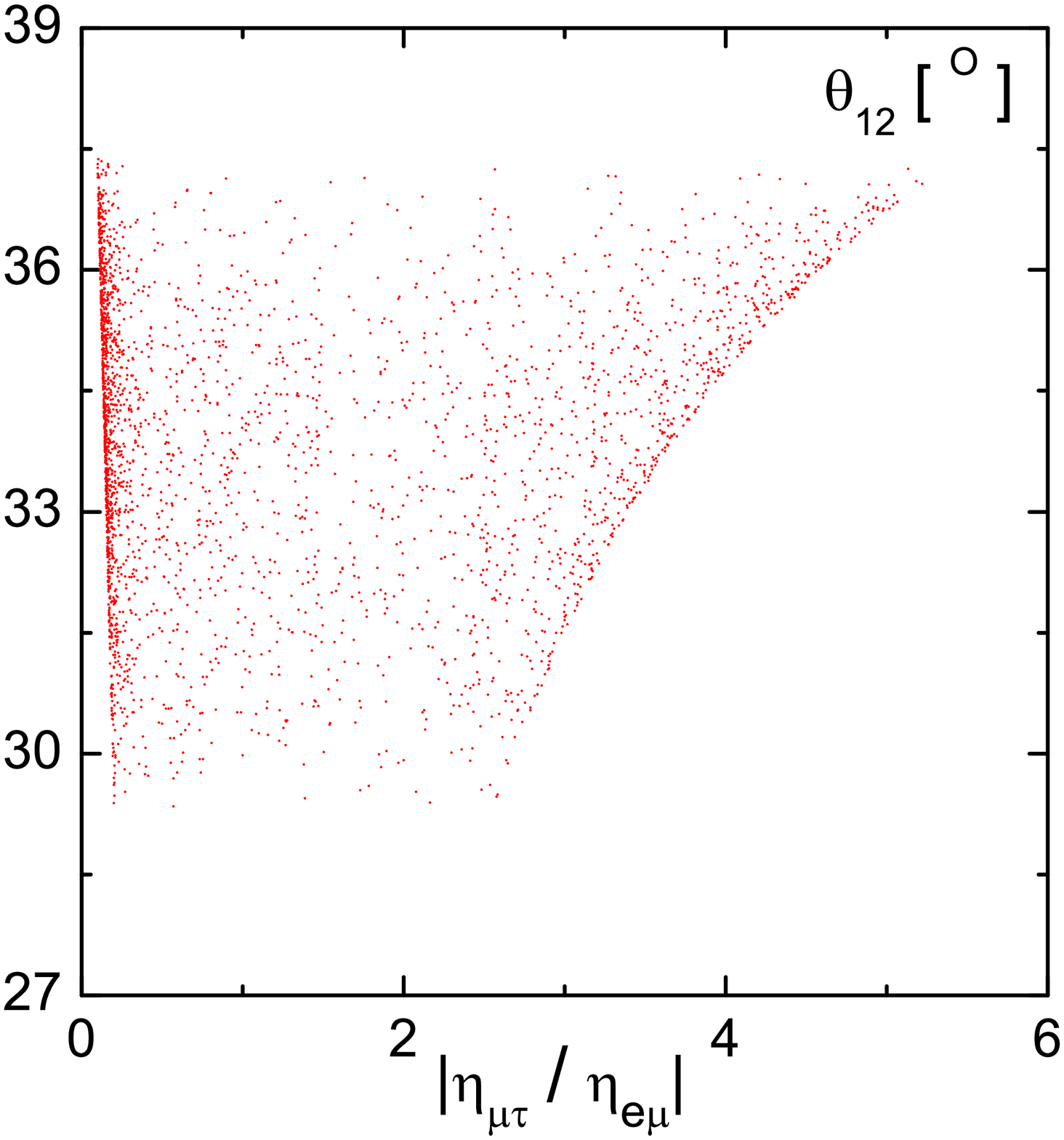}
\includegraphics[width=6.0cm,bb=30 120 680 770]{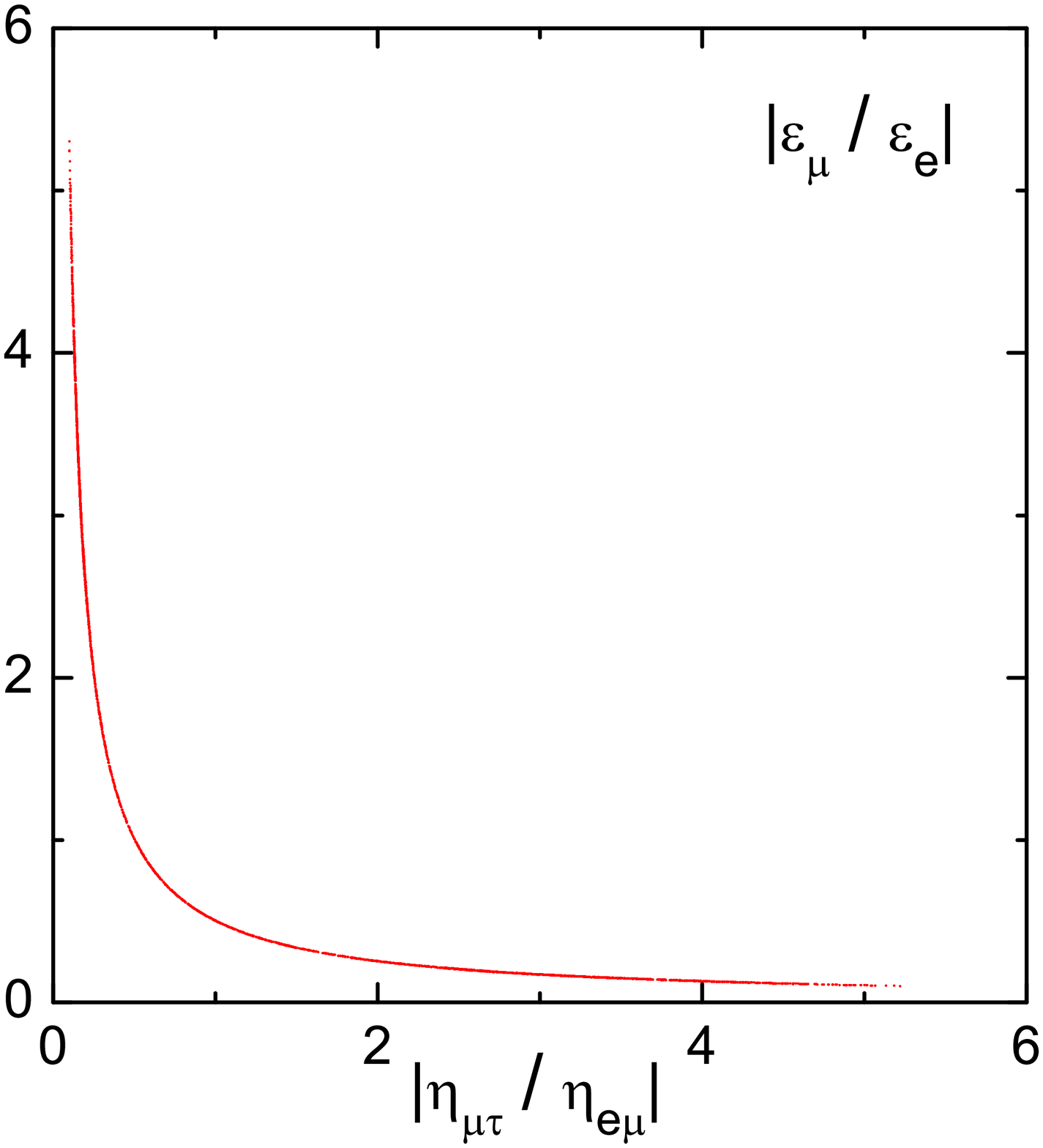}
\includegraphics[width=6.0cm,bb=30 120 680 770]{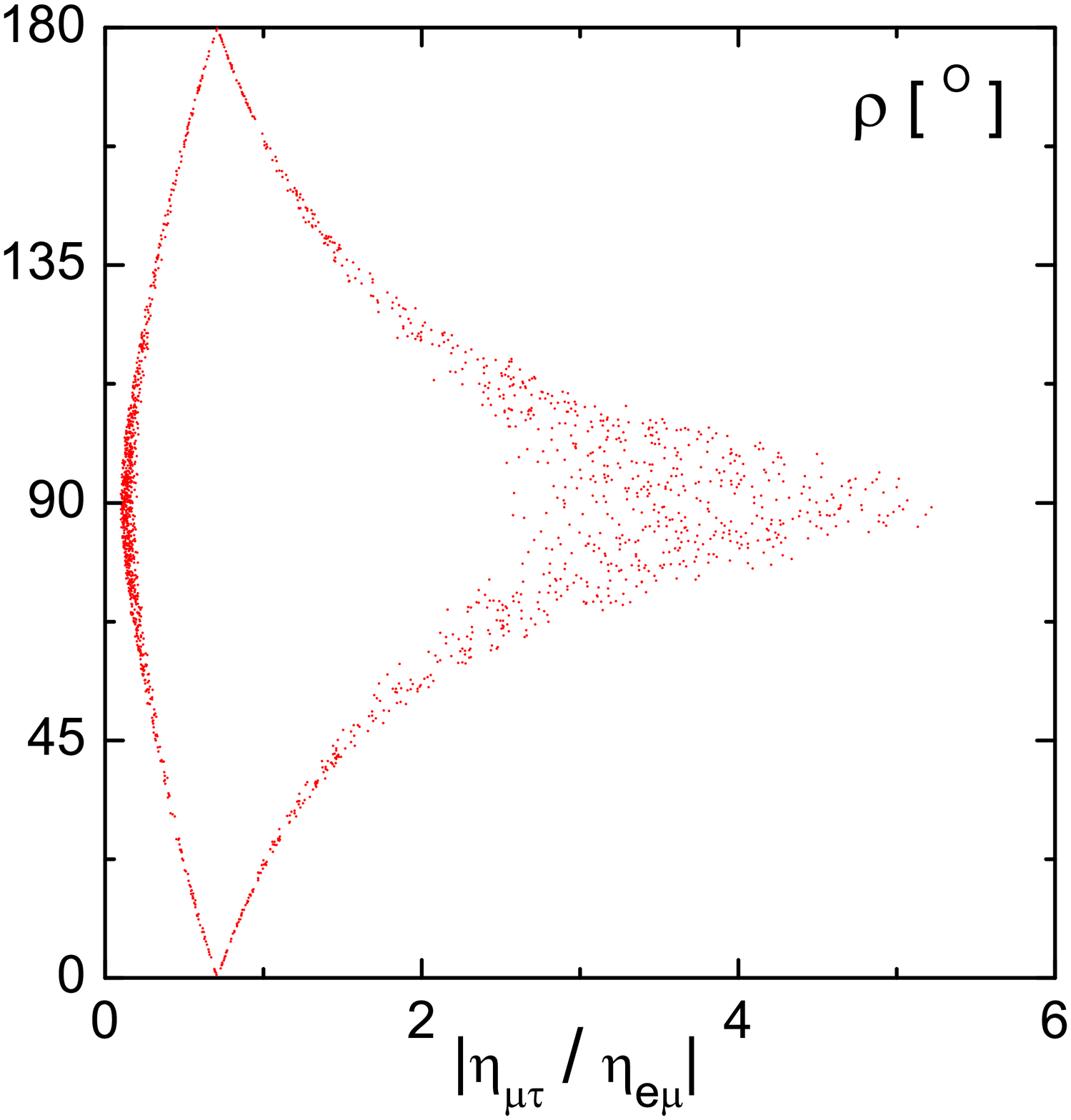}
\vspace{1cm} \caption{\label{fig:fig1} The parameter space of the
neutrino mixing angle $\theta^{}_{12}$, the Majorana phase $\rho$,
and the ratios $(\varepsilon^{}_\mu/\varepsilon^{}_e, \eta^{}_{\mu
\tau}/\eta^{}_{e\mu})$.} \vspace{-0.5cm}
\end{center}
\end{figure}

It is worthwhile to note that the ratios
$\varepsilon^{}_\mu/\varepsilon^{}_e$ and $y^{}_\mu/y^{}_e$ are of
great importance in our model. For a given mass $M$ (or equivalently
$\omega \equiv v/M$) and also the mass scale of light neutrinos
$\omega y^{}_e |\varepsilon^{}_e|$, all the other observables are
determined by these ratios. One can always appropriately rescale
$\varepsilon^{}_\alpha$ and $y^{}_\alpha$ to get the desired masses
$m^{}_i$, keeping the ratios $\varepsilon^{}_\mu/\varepsilon^{}_e$
and $y^{}_\mu/y^{}_e$ unchanged. Therefore, we shall mainly
concentrate on the relative ratios among model parameters in the
following discussions.

In FIG.~\ref{fig:fig1}, we show the allowed parameter space of the
model within 3$\sigma$ C.L. In our numerical analysis, we take
$M=1~{\rm TeV}$ for example, as well as the values of neutrino
masses and mixing angles from Ref.~\cite{Schwetz:2008er}. We also
include the experimental constraints on the non-unitary effects
coming from universality tests of weak interactions, rare leptonic
decays, invisible width of the $Z$-boson and neutrino oscillation
data \cite{Antusch:2008tz}. Technologically, we randomly choose the
values of ($\varepsilon^{}_e,\varepsilon^{}_\mu,y^{}_e,y^{}_\mu$)
and their corresponding phases, while the data sample reproducing
all the neutrino masses and mixing angles within 3$\sigma$
confidence ranges will be kept. The absolute scales of
$\varepsilon^{}_\alpha$ and $y^{}_\alpha$ (for $\alpha = e, \mu$)
are also checked to be consistent with the experimental bounds on
$\eta$. Some comments are in order:
\begin{itemize}
\item From the uppermost plot, one can see that the ratio
$|\eta^{}_{\mu\tau}/\eta^{}_{e\mu}| = |y^{}_\mu/y^{}_e|$ is strictly
constrained by $\theta^{}_{12}$ at large values, and the maximal
value of the ratio is close to 5. In this case, the current bound
$|\eta_{e\mu}|<6\times10^{-5}$ suggests a rather stringent bound
$|\eta_{\mu\tau}|\lesssim 3\times 10^{-4}$. However, as pointed out
in Ref.~\cite{Antusch:2008tz}, in the case with $M < \Lambda_{\rm
EW}$ but above a few GeV, the severe constraints from $Z$ and $W$
decays do not apply since the unitarity is restored. In this special
case the upper bound $|\eta_{e\mu}|<9\times10^{-4}$ indicates that
there is no general constraint on $|\eta_{\mu\tau}|$.

\item The correlation between $|\varepsilon_\mu/\varepsilon_e|$ and
$|\eta_{\mu\tau}/\eta_{e\mu}|$ is illustrated in the middle plot.
The approximate relation
$|\varepsilon_\mu/\varepsilon_e|\times|\eta_{e\mu}/\eta_{\mu\tau}|
\approx 0.5$ reflects the fact that $|A|$ and $2|C|$  in
Eq.~\eqref{eq:mnu3} are comparable in magnitude to ensure the
experimental result $m^{}_1 \simeq m^{}_2$. To generate a large but
non-maximal mixing angle $\theta^{}_{12}$, the relation $|B| \ll
|A|$ must hold, which is only possible if the relative phase between
$A$ and $C$ (or equivalently $\varepsilon^{}_e$ and
$\varepsilon^{}_\mu$) is around $180^\circ$. However, the exact
identity $|A| = 2|C|$ is forbidden due to the mixing angle $\theta$
in Eq.~\eqref{eq:theta}, which exhibits the strict experimental
constraints on model parameters.

\item In the last plot, we show the allowed ranges of the Majorana
phase $\rho$ versus the ratio $|\eta_{\mu\tau}/\eta_{e\mu}|$. As we
have discussed, the requirement $A \simeq -2C$ also sets a strong
constraint on the relative phase between $\varepsilon_e$ and
$\varepsilon_\mu$. For a sizable $\eta_{\mu\tau}$, which is
phenomenologically interesting, $\rho \sim 90^\circ$ can be read
from the plot. For intermediate values of $\eta_{\mu\tau}$, we can see
that $\rho$ deviates significantly from $90^\circ$. In principle,
the whole range $[0, 180^\circ]$ of $\rho$ is allowed.
\end{itemize}
As a consequence of the minimal number of model parameters, the
mixing angle $\theta^{}_{12}$, the Majorana phase $\rho$ and the
non-unitarity parameters $|\eta^{}_{\alpha \beta}|$ are now related
to one another. Furthermore, the introduction of non-vanishing
$\theta^{}_{13}$ and CP-violating phase is straightforward by
relaxing the $\mu$-$\tau$ symmetry.

\section{Further Discussions}\label{sec:LHC}

Now we shall address the possibly intriguing low-energy
phenomena of our model, as well as the signals of the heavy
neutrinos at the LHC.

\paragraph{Rare LFV decays mediated by the heavy neutrino.}
The heavy neutrinos entering into the charged-current interaction
will contribute to the LFV decays of charged leptons. For instance,
the decay channel $\ell^{}_\alpha \rightarrow \ell_\beta \gamma$ can
be mediated by heavy neutrinos \cite{Ilakovac:1994kj}. In the
standard type-I seesaw scenario or MSM, one has approximately
$(M^{}_{\rm D}/M)^2 = {\cal O}(m M^{-1})$, and therefore ${\rm
BR}\left({\ell_\alpha \rightarrow \ell_\beta \gamma}\right) \propto
{\cal O} (m^2)$ indicates a strong suppression of LFV decay rates,
where $m$ and $M$ denote respectively the scales of light and heavy
neutrino masses. However, in our model, one can have sizeable
$M^{}_{\rm D}/M$ without facing the difficulty of neutrino mass
generation since they are suppressed by the perturbations
$\varepsilon^{}_\alpha$. Thus appreciable LFV rates could be
obtained even for strictly massless light neutrinos
\cite{Bernabeu:1987gr}.

\begin{figure}[t]
\begin{center}\vspace{-0.7cm}
\includegraphics[width=6.cm,bb=30 100 680 750]{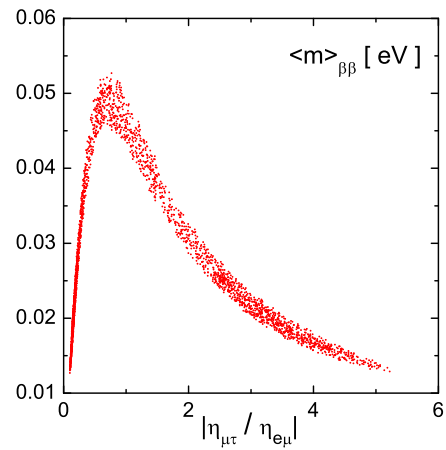}
\vspace{0.9cm} \caption{\label{fig:fig2} The allowed region of the
effective neutrino mass $\langle m\rangle^{}_{\beta \beta}$ versus
the ratio $|\eta^{}_{\mu\tau}/\eta^{}_{e\mu}|$.} \vspace{-0.5cm}
\end{center}
\end{figure}

\paragraph{Neutrinoless double beta decays ($0\nu2\beta$).} The
light Majorana neutrinos contribute to $0\nu2\beta$ decays, which
serve as the unique tool to discriminate between Majorana and Dirac
nature of massive neutrinos. The relevant quantity is the effective
neutrino mass $\langle m\rangle^{}_{\beta \beta} \equiv |m^{}_1
U^2_{e1} + m^{}_2 U^2_{e2} + m^{}_3 U^2_{e3}|$, which in our model
with $m^{}_3 = 0$ can be evaluated as $\langle m\rangle^{}_{\beta
\beta} = |m^{}_1 U^2_{e1} + m^{}_2 U^2_{e2}|$. Because of $m^{}_2
> m^{}_1 = \sqrt{|\Delta m^2_{31}|} \approx 0.05~{\rm eV}$, the two
terms in the $\langle m\rangle^{}_{\beta \beta}$ are comparable in
magnitude. In this case, the Majorana phase plays a key role in
determining $\langle m\rangle^{}_{\beta \beta}$. For example, if
$\rho$ is far away from $90^\circ$, the contributions from these two
terms should be added constructively, and one can then expect a
large value $\langle m\rangle^{}_{\beta \beta}\sim 0.05 ~ {\rm eV}$.
In FIG.~\ref{fig:fig2} we have shown the allowed region of $\langle
m\rangle^{}_{\beta \beta}$. The result is in agreement with that in
FIG.~\ref{fig:fig1}. It is worth noting that heavy Majorana
neutrinos are nearly degenerate in mass, i.e. they form a
pseudo-Dirac neutrino, so their contributions to $\langle
m\rangle^{}_{\beta \beta}$ can be neglected. Interestingly, the
next-generation $0\nu2\beta$ decay experiments are expected to probe
$\langle m\rangle^{}_{\beta \beta}$ with the accuracy of $10-50~{\rm
meV}$, so our model can be tested experimentally in the near future.

\paragraph{Search for the tri-lepton signals at the LHC.}
As shown in Eq.~\eqref{eq:Lcc}, the heavy singlet $P$ couples to the
gauge sector of the SM, and thus if kinematically accessible, could
be produced at hadron colliders. In the case $M > M^{}_H$ (where
$M^{}_H$ denotes the Higgs mass), the heavy neutrino can decay in
the channels $P \to \ell^+ + W^-$, $P \to \bar{\nu} + Z$, and $P \to
\bar{\nu} + H$. Now that the heavy neutrinos form a pseudo-Dirac
particle, we shall focus our attention on the
lepton-number-conserving processes induced by it. For example, one
very interesting and prospective channel is the production of three
charged leptons with missing energy \cite{delAguila:2008hw}, i.e.
$pp\to \ell_{\alpha}^{\pm} \ell_{\beta}^{\pm}\ell_{\gamma}^{\mp}
\nu(\bar\nu)+\text{jets}$. Another possible one is the pair
production of charged leptons in different flavors and without
missing energy, i.e. $pp\to \ell_{\alpha}^{\pm}\ell_{\beta}^{\mp} +
\mbox{jets}$. However, it is difficult to make significant
observation in this channel at the LHC due to the large SM
background \cite{Aguila:2007em}.

\section{summary}\label{sec:summary}

In this work, we have considered a novel minimal seesaw model, which
is quite natural to bring down the typical seesaw mechanism to the
electroweak scale. Our model is minimal among the fermionic seesaw
scenarios in the sense of particle content, since only two
right-handed neutrinos are introduced. However, quite different from
the traditional MSM, the heavy right-handed neutrinos in the current
consideration is located around the TeV scale, and possess sizable
Yukawa couplings, which make them observable, in particular at the
LHC. Furthermore, light neutrino masses are protected by a global
$U(1)$ symmetry, and thus free from radiative instability. Since the
model contains only two heavy singlet neutrinos, it is very
predictive compared to the other seesaw mechanisms. We have shown
that, in the assumption of a $\mu \leftrightarrow \tau$ symmetry,
the tri-bimaximal pattern of lepton flavor mixing can be reproduced
and only the inverted neutrino mass hierarchy is allowed. The light
neutrino masses, the Majorana phase and non-unitarity parameters are
connected in our model, and their non-trivial correlations could be
tested through a combined analysis of the future long-baseline
neutrino oscillation experiments, the rare LFV decays, the
$0\nu2\beta$ decays and the collider signals at the LHC.

\begin{acknowledgments}

One of the authors (S.Z.) is indebted to G.G. Raffelt for warm
hospitality at Max-Planck-Institut f\"{u}r Physik. This work was
supported by the Royal Institute of Technology (KTH), project
no.~SII-56510 (H.Z.), and the Alexander von Humboldt Foundation
(S.Z.).

\end{acknowledgments}

\bibliography{bib}

\begin{thebibliography}{32}
\expandafter\ifx\csname natexlab\endcsname\relax\def\natexlab#1{#1}\fi
\expandafter\ifx\csname bibnamefont\endcsname\relax
  \def\bibnamefont#1{#1}\fi
\expandafter\ifx\csname bibfnamefont\endcsname\relax
  \def\bibfnamefont#1{#1}\fi
\expandafter\ifx\csname citenamefont\endcsname\relax
  \def\citenamefont#1{#1}\fi
\expandafter\ifx\csname url\endcsname\relax
  \def\url#1{\texttt{#1}}\fi
\expandafter\ifx\csname urlprefix\endcsname\relax\def\urlprefix{URL }\fi
\providecommand{\bibinfo}[2]{#2}
\providecommand{\eprint}[2][]{\url{#2}}

\bibitem[{\citenamefont{Minkowski}(1977)}]{Minkowski:1977sc}
\bibinfo{author}{\bibfnamefont{P.}~\bibnamefont{Minkowski}},
  \bibinfo{journal}{Phys. Lett.} \textbf{\bibinfo{volume}{B67}},
  \bibinfo{pages}{421} (\bibinfo{year}{1977}).

\bibitem[{\citenamefont{Yanagida}(1979)}]{Yanagida:1979as}
\bibinfo{author}{\bibfnamefont{T.}~\bibnamefont{Yanagida}}, in
  \emph{\bibinfo{booktitle}{Proc. Workshop on the Baryon Number of the Universe
  and Unified Theories}}, edited by
  \bibinfo{editor}{\bibfnamefont{O.}~\bibnamefont{Sawada}} \bibnamefont{and}
  \bibinfo{editor}{\bibfnamefont{A.}~\bibnamefont{Sugamoto}}
  (\bibinfo{year}{1979}), p.~\bibinfo{pages}{95}.

\bibitem[{\citenamefont{Gell-Mann et~al.}(1979)\citenamefont{Gell-Mann, Ramond,
  and Slansky}}]{GellMann:1980vs}
\bibinfo{author}{\bibfnamefont{M.}~\bibnamefont{Gell-Mann}},
  \bibinfo{author}{\bibfnamefont{P.}~\bibnamefont{Ramond}}, \bibnamefont{and}
  \bibinfo{author}{\bibfnamefont{R.}~\bibnamefont{Slansky}}, in
  \emph{\bibinfo{booktitle}{Supergravity}}, edited by
  \bibinfo{editor}{\bibfnamefont{P.}~\bibnamefont{van Nieuwenhuizen}}
  \bibnamefont{and} \bibinfo{editor}{\bibfnamefont{D.}~\bibnamefont{Freedman}}
  (\bibinfo{year}{1979}), p. \bibinfo{pages}{315}.

\bibitem[{\citenamefont{Mohapatra and Senjanovi{\'c}}(1980)}]{Mohapatra:1979ia}
\bibinfo{author}{\bibfnamefont{R.~N.} \bibnamefont{Mohapatra}}
  \bibnamefont{and}
  \bibinfo{author}{\bibfnamefont{G.}~\bibnamefont{Senjanovi{\'c}}},
  \bibinfo{journal}{Phys. Rev. Lett.} \textbf{\bibinfo{volume}{44}},
  \bibinfo{pages}{912} (\bibinfo{year}{1980}).

\bibitem[{\citenamefont{Pilaftsis}(1992)}]{Pilaftsis:1991ug}
\bibinfo{author}{\bibfnamefont{A.}~\bibnamefont{Pilaftsis}},
  \bibinfo{journal}{Z. Phys.} \textbf{\bibinfo{volume}{C55}},
  \bibinfo{pages}{275} (\bibinfo{year}{1992}), \eprint{hep-ph/9901206}.

\bibitem[{\citenamefont{Kersten and Smirnov}(2007)}]{Kersten:2007vk}
\bibinfo{author}{\bibfnamefont{J.}~\bibnamefont{Kersten}} \bibnamefont{and}
  \bibinfo{author}{\bibfnamefont{A.~Y.} \bibnamefont{Smirnov}},
  \bibinfo{journal}{Phys. Rev.} \textbf{\bibinfo{volume}{D76}},
  \bibinfo{pages}{073005} (\bibinfo{year}{2007}), \eprint{arXiv:0705.3221}.

\bibitem[{\citenamefont{Mohapatra and Valle}(1986)}]{Mohapatra:1986bd}
\bibinfo{author}{\bibfnamefont{R.~N.} \bibnamefont{Mohapatra}}
  \bibnamefont{and} \bibinfo{author}{\bibfnamefont{J.~W.~F.}
  \bibnamefont{Valle}}, \bibinfo{journal}{Phys. Rev.}
  \textbf{\bibinfo{volume}{D34}}, \bibinfo{pages}{1642} (\bibinfo{year}{1986}).

\bibitem[{\citenamefont{Malinsk{\'y}
  et~al.}(2009{\natexlab{a}})\citenamefont{Malinsk{\'y}, Ohlsson, Xing, and
  Zhang}}]{Malinsky:2009df}
\bibinfo{author}{\bibfnamefont{M.}~\bibnamefont{Malinsk{\'y}}},
  \bibinfo{author}{\bibfnamefont{T.}~\bibnamefont{Ohlsson}},
  \bibinfo{author}{\bibfnamefont{Z.-z.} \bibnamefont{Xing}}, \bibnamefont{and}
  \bibinfo{author}{\bibfnamefont{H.}~\bibnamefont{Zhang}},
  \bibinfo{journal}{Phys. Lett.} \textbf{\bibinfo{volume}{B679}},
  \bibinfo{pages}{242} (\bibinfo{year}{2009}{\natexlab{a}}),
  \eprint{arXiv:0905.2889}.

\bibitem[{\citenamefont{Xing and Zhou}(2009)}]{Xing:2009hx}
\bibinfo{author}{\bibfnamefont{Z.-z.} \bibnamefont{Xing}} \bibnamefont{and}
  \bibinfo{author}{\bibfnamefont{S.}~\bibnamefont{Zhou}},
  \bibinfo{journal}{Phys. Lett.} \textbf{\bibinfo{volume}{B679}},
  \bibinfo{pages}{249} (\bibinfo{year}{2009}), \eprint{arXiv:0906.1757}.

\bibitem[{\citenamefont{Yamanaka et~al.}(1982)\citenamefont{Yamanaka, Sugawara,
  and Pakvasa}}]{Yamanaka:1981pa}
\bibinfo{author}{\bibfnamefont{Y.}~\bibnamefont{Yamanaka}},
  \bibinfo{author}{\bibfnamefont{H.}~\bibnamefont{Sugawara}}, \bibnamefont{and}
  \bibinfo{author}{\bibfnamefont{S.}~\bibnamefont{Pakvasa}},
  \bibinfo{journal}{Phys. Rev.} \textbf{\bibinfo{volume}{D25}},
  \bibinfo{pages}{1895} (\bibinfo{year}{1982}).

\bibitem[{\citenamefont{Zhang}(2007)}]{Zhang:2006fv}
\bibinfo{author}{\bibfnamefont{H.}~\bibnamefont{Zhang}},
  \bibinfo{journal}{Phys. Lett.} \textbf{\bibinfo{volume}{B655}},
  \bibinfo{pages}{132} (\bibinfo{year}{2007}), \eprint{hep-ph/0612214}.

\bibitem[{\citenamefont{Frampton et~al.}(2002)\citenamefont{Frampton, Glashow,
  and Yanagida}}]{Frampton:2002qc}
\bibinfo{author}{\bibfnamefont{P.~H.} \bibnamefont{Frampton}},
  \bibinfo{author}{\bibfnamefont{S.~L.} \bibnamefont{Glashow}},
  \bibnamefont{and} \bibinfo{author}{\bibfnamefont{T.}~\bibnamefont{Yanagida}},
  \bibinfo{journal}{Phys. Lett.} \textbf{\bibinfo{volume}{B548}},
  \bibinfo{pages}{119} (\bibinfo{year}{2002}), \eprint{hep-ph/0208157}.

\bibitem[{\citenamefont{Guo et~al.}(2007)\citenamefont{Guo, Xing, and
  Zhou}}]{Guo:2006qa}
\bibinfo{author}{\bibfnamefont{W.-l.} \bibnamefont{Guo}},
  \bibinfo{author}{\bibfnamefont{Z.-z.} \bibnamefont{Xing}}, \bibnamefont{and}
  \bibinfo{author}{\bibfnamefont{S.}~\bibnamefont{Zhou}},
  \bibinfo{journal}{Int. J. Mod. Phys.} \textbf{\bibinfo{volume}{E16}},
  \bibinfo{pages}{1} (\bibinfo{year}{2007}), \eprint{hep-ph/0612033}.

\bibitem[{\citenamefont{Bilenky and Petcov}(1987)}]{Bilenky:1987ty}
\bibinfo{author}{\bibfnamefont{S.~M.} \bibnamefont{Bilenky}} \bibnamefont{and}
  \bibinfo{author}{\bibfnamefont{S.~T.} \bibnamefont{Petcov}},
  \bibinfo{journal}{Rev. Mod. Phys.} \textbf{\bibinfo{volume}{59}},
  \bibinfo{pages}{671} (\bibinfo{year}{1987}).

\bibitem[{\citenamefont{Schechter and Valle}(1980)}]{Schechter:1980gr}
\bibinfo{author}{\bibfnamefont{J.}~\bibnamefont{Schechter}} \bibnamefont{and}
  \bibinfo{author}{\bibfnamefont{J.~W.~F.} \bibnamefont{Valle}},
  \bibinfo{journal}{Phys. Rev.} \textbf{\bibinfo{volume}{D22}},
  \bibinfo{pages}{2227} (\bibinfo{year}{1980}).

\bibitem[{\citenamefont{Lazarides et~al.}(1981)\citenamefont{Lazarides, Shafi,
  and Wetterich}}]{Lazarides:1980nt}
\bibinfo{author}{\bibfnamefont{G.}~\bibnamefont{Lazarides}},
  \bibinfo{author}{\bibfnamefont{Q.}~\bibnamefont{Shafi}}, \bibnamefont{and}
  \bibinfo{author}{\bibfnamefont{C.}~\bibnamefont{Wetterich}},
  \bibinfo{journal}{Nucl. Phys.} \textbf{\bibinfo{volume}{B181}},
  \bibinfo{pages}{287} (\bibinfo{year}{1981}).

\bibitem[{\citenamefont{Mohapatra and Senjanovi{\'c}}(1981)}]{Mohapatra:1980yp}
\bibinfo{author}{\bibfnamefont{R.~N.} \bibnamefont{Mohapatra}}
  \bibnamefont{and}
  \bibinfo{author}{\bibfnamefont{G.}~\bibnamefont{Senjanovi{\'c}}},
  \bibinfo{journal}{Phys. Rev.} \textbf{\bibinfo{volume}{D23}},
  \bibinfo{pages}{165} (\bibinfo{year}{1981}).

\bibitem[{\citenamefont{Malinsk{\'y}
  et~al.}(2009{\natexlab{b}})\citenamefont{Malinsk{\'y}, Ohlsson, and
  Zhang}}]{Malinsky:2008qn}
\bibinfo{author}{\bibfnamefont{M.}~\bibnamefont{Malinsk{\'y}}},
  \bibinfo{author}{\bibfnamefont{T.}~\bibnamefont{Ohlsson}}, \bibnamefont{and}
  \bibinfo{author}{\bibfnamefont{H.}~\bibnamefont{Zhang}},
  \bibinfo{journal}{Phys. Rev.} \textbf{\bibinfo{volume}{D79}},
  \bibinfo{pages}{011301(R)} (\bibinfo{year}{2009}{\natexlab{b}}),
  \eprint{arXiv:0811.3346}.

\bibitem[{\citenamefont{Foot et~al.}(1989)\citenamefont{Foot, Lew, He, and
  Joshi}}]{Foot:1988aq}
\bibinfo{author}{\bibfnamefont{R.}~\bibnamefont{Foot}},
  \bibinfo{author}{\bibfnamefont{H.}~\bibnamefont{Lew}},
  \bibinfo{author}{\bibfnamefont{X.~G.} \bibnamefont{He}}, \bibnamefont{and}
  \bibinfo{author}{\bibfnamefont{G.~C.} \bibnamefont{Joshi}},
  \bibinfo{journal}{Z. Phys.} \textbf{\bibinfo{volume}{C44}},
  \bibinfo{pages}{441} (\bibinfo{year}{1989}).

\bibitem[{\citenamefont{Malinsk{\'y}
  et~al.}(2009{\natexlab{c}})\citenamefont{Malinsk{\'y}, Ohlsson, and
  Zhang}}]{Malinsky:2009gw}
\bibinfo{author}{\bibfnamefont{M.}~\bibnamefont{Malinsk{\'y}}},
  \bibinfo{author}{\bibfnamefont{T.}~\bibnamefont{Ohlsson}}, \bibnamefont{and}
  \bibinfo{author}{\bibfnamefont{H.}~\bibnamefont{Zhang}}
  (\bibinfo{year}{2009}{\natexlab{c}}), \eprint{arXiv:0903.1961}.

\bibitem[{\citenamefont{Pilaftsis and Underwood}(2005)}]{Pilaftsis:2005rv}
\bibinfo{author}{\bibfnamefont{A.}~\bibnamefont{Pilaftsis}} \bibnamefont{and}
  \bibinfo{author}{\bibfnamefont{T.~E.~J.} \bibnamefont{Underwood}},
  \bibinfo{journal}{Phys. Rev.} \textbf{\bibinfo{volume}{D72}},
  \bibinfo{pages}{113001} (\bibinfo{year}{2005}), \eprint{hep-ph/0506107}.

\bibitem[{\citenamefont{Malinsk{\'y} et~al.}(2005)\citenamefont{Malinsk{\'y},
  Romao, and Valle}}]{Malinsky:2005bi}
\bibinfo{author}{\bibfnamefont{M.}~\bibnamefont{Malinsk{\'y}}},
  \bibinfo{author}{\bibfnamefont{J.~C.} \bibnamefont{Romao}}, \bibnamefont{and}
  \bibinfo{author}{\bibfnamefont{J.~W.~F.} \bibnamefont{Valle}},
  \bibinfo{journal}{Phys. Rev. Lett.} \textbf{\bibinfo{volume}{95}},
  \bibinfo{pages}{161801} (\bibinfo{year}{2005}), \eprint{hep-ph/0506296}.

\bibitem[{\citenamefont{Schechter and Valle}(1982)}]{Schechter:1981cv}
\bibinfo{author}{\bibfnamefont{J.}~\bibnamefont{Schechter}} \bibnamefont{and}
  \bibinfo{author}{\bibfnamefont{J.~W.~F.} \bibnamefont{Valle}},
  \bibinfo{journal}{Phys. Rev.} \textbf{\bibinfo{volume}{D25}},
  \bibinfo{pages}{774} (\bibinfo{year}{1982}).

\bibitem[{\citenamefont{Harrison et~al.}(2002)\citenamefont{Harrison, Perkins,
  and Scott}}]{Harrison:2002er}
\bibinfo{author}{\bibfnamefont{P.~F.} \bibnamefont{Harrison}},
  \bibinfo{author}{\bibfnamefont{D.~H.} \bibnamefont{Perkins}},
  \bibnamefont{and} \bibinfo{author}{\bibfnamefont{W.~G.} \bibnamefont{Scott}},
  \bibinfo{journal}{Phys. Lett.} \textbf{\bibinfo{volume}{B530}},
  \bibinfo{pages}{167} (\bibinfo{year}{2002}), \eprint{hep-ph/0202074}.

\bibitem[{\citenamefont{Xing}(2002)}]{Xing:2002sw}
\bibinfo{author}{\bibfnamefont{Z.-z.} \bibnamefont{Xing}},
  \bibinfo{journal}{Phys. Lett.} \textbf{\bibinfo{volume}{B533}},
  \bibinfo{pages}{85} (\bibinfo{year}{2002}), \eprint{hep-ph/0204049}.

\bibitem[{\citenamefont{Altarelli and Meloni}(2009)}]{Altarelli:2008yr}
\bibinfo{author}{\bibfnamefont{G.}~\bibnamefont{Altarelli}} \bibnamefont{and}
  \bibinfo{author}{\bibfnamefont{D.}~\bibnamefont{Meloni}},
  \bibinfo{journal}{Nucl. Phys.} \textbf{\bibinfo{volume}{B809}},
  \bibinfo{pages}{158} (\bibinfo{year}{2009}), \eprint{arXiv:0809.1041}.

\bibitem[{\citenamefont{Schwetz et~al.}(2008)\citenamefont{Schwetz,
  T{\'o}rtola, and Valle}}]{Schwetz:2008er}
\bibinfo{author}{\bibfnamefont{T.}~\bibnamefont{Schwetz}},
  \bibinfo{author}{\bibfnamefont{M.~A.} \bibnamefont{T{\'o}rtola}},
  \bibnamefont{and} \bibinfo{author}{\bibfnamefont{J.~W.~F.}
  \bibnamefont{Valle}}, \bibinfo{journal}{New J. Phys.}
  \textbf{\bibinfo{volume}{10}}, \bibinfo{pages}{113011}
  (\bibinfo{year}{2008}), \eprint{arXiv:0808.2016}.

\bibitem[{\citenamefont{Antusch et~al.}(2009)\citenamefont{Antusch, Baumann,
  and Fern{\'a}ndez-Mart{\'i}nez}}]{Antusch:2008tz}
\bibinfo{author}{\bibfnamefont{S.}~\bibnamefont{Antusch}},
  \bibinfo{author}{\bibfnamefont{J.~P.} \bibnamefont{Baumann}},
  \bibnamefont{and}
  \bibinfo{author}{\bibfnamefont{E.}~\bibnamefont{Fern{\'a}ndez-Mart{\'i}nez}},
  \bibinfo{journal}{Nucl. Phys.} \textbf{\bibinfo{volume}{B810}},
  \bibinfo{pages}{369} (\bibinfo{year}{2009}), \eprint{arXiv:0807.1003}.

\bibitem[{\citenamefont{Ilakovac and Pilaftsis}(1995)}]{Ilakovac:1994kj}
\bibinfo{author}{\bibfnamefont{A.}~\bibnamefont{Ilakovac}} \bibnamefont{and}
  \bibinfo{author}{\bibfnamefont{A.}~\bibnamefont{Pilaftsis}},
  \bibinfo{journal}{Nucl. Phys.} \textbf{\bibinfo{volume}{B437}},
  \bibinfo{pages}{491} (\bibinfo{year}{1995}), \eprint{hep-ph/9403398}.

\bibitem[{\citenamefont{Bernab{\'e}u et~al.}(1987)\citenamefont{Bernab{\'e}u,
  Santamaria, Vidal, Mendez, and Valle}}]{Bernabeu:1987gr}
\bibinfo{author}{\bibfnamefont{J.}~\bibnamefont{Bernab{\'e}u}},
  \bibinfo{author}{\bibfnamefont{A.}~\bibnamefont{Santamaria}},
  \bibinfo{author}{\bibfnamefont{J.}~\bibnamefont{Vidal}},
  \bibinfo{author}{\bibfnamefont{A.}~\bibnamefont{Mendez}}, \bibnamefont{and}
  \bibinfo{author}{\bibfnamefont{J.~W.~F.} \bibnamefont{Valle}},
  \bibinfo{journal}{Phys. Lett.} \textbf{\bibinfo{volume}{B187}},
  \bibinfo{pages}{303} (\bibinfo{year}{1987}).

\bibitem[{\citenamefont{del Aguila and
  Aguilar-Saavedra}(2009)}]{delAguila:2008hw}
\bibinfo{author}{\bibfnamefont{F.}~\bibnamefont{del Aguila}} \bibnamefont{and}
  \bibinfo{author}{\bibfnamefont{J.~A.} \bibnamefont{Aguilar-Saavedra}},
  \bibinfo{journal}{Phys. Lett.} \textbf{\bibinfo{volume}{B672}},
  \bibinfo{pages}{158} (\bibinfo{year}{2009}), \eprint{arXiv:0809.2096}.

\bibitem[{\citenamefont{del Aguila et~al.}(2007)\citenamefont{del Aguila,
  Aguilar-Saavedra, and Pittau}}]{Aguila:2007em}
\bibinfo{author}{\bibfnamefont{F.}~\bibnamefont{del Aguila}},
  \bibinfo{author}{\bibfnamefont{J.~A.} \bibnamefont{Aguilar-Saavedra}},
  \bibnamefont{and} \bibinfo{author}{\bibfnamefont{R.}~\bibnamefont{Pittau}},
  \bibinfo{journal}{JHEP} \textbf{\bibinfo{volume}{10}}, \bibinfo{pages}{047}
  (\bibinfo{year}{2007}), \eprint{hep-ph/0703261}.

\end{thebibliography}

\end{document}